\documentclass[final]{aipproc}
\layoutstyle{6x9}
\begin{document}
\title{Searching for Planets with White Dwarf Pulsations: Spurious Detections}

\classification{97.10.Sj, 97.20.Rp, 97.82.Fs}
\keywords      {Exoplanets, Pulsation Timing, O-C, White Dwarfs}

\author{James Dalessio}{
  address={University of Delaware, Department of Physics and Astronomy, 217 Sharp Lab, Newark~DE~19716, USA},
  altaddress={Delaware Asteroseismology Research Center}
}

\author{Judith L.\ Provencal}{
  address={University of Delaware, Department of Physics and Astronomy, 217 Sharp Lab, Newark~DE~19716, USA},
  altaddress={Delaware Asteroseismology Research Center}
}

\author{Harry S.\ Shipman}{
  address={University of Delaware, Department of Physics and Astronomy, 217 Sharp Lab, Newark~DE~19716, USA},
  altaddress={Delaware Asteroseismology Research Center}
}

\begin{abstract}
We present 13 years of pulsation timing measurements of the DBV white dwarf EC\,20058$-$5234.  Each of the four O$-$C diagrams mimic the sinusoidal behavior typically attributed to a planet + WD system.  However, the amplitude and phase of the O$-$C variations are inconsistent with each other.  We discuss the impact of this result on timing based WD planet searches.
\end{abstract}

\maketitle

\section{Introduction}
EC\,20058$-$5234 has been the target of a long term campaign to measure the neutrino production rate in hot DBV WDs. These measurements also allow us to probe for a planetary system using the pulsation timing method, see \cite{2005ASPC..335....3S}. A planetary companion will cause a wobble of the WD around the system barycenter, resulting in variations between the expected and observed arrival times of the WD pulsations (O$-$C method). This will cause an identical sinusoidal variation in each frequency's O$-$C diagram. This method has resulted in a planet detection around an sdB star, see \cite{2007Natur.449..189S} and planetary candidates and exclusions around WDs, see \cite{2008ApJ...676..573M}.

\section{The O$-$C Diagram of EC\,20058$-$5234}
\begin{figure} \label{OC}
  \includegraphics[width=6in]{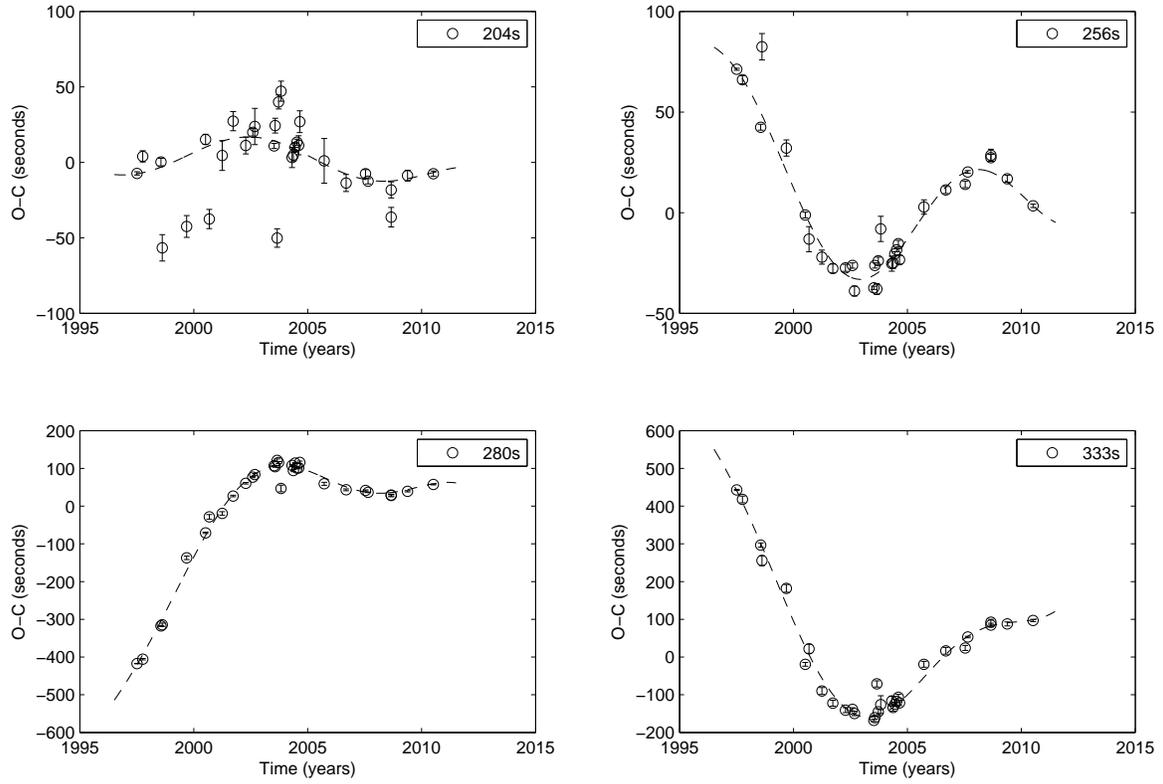}
  \caption{O-C diagram and model fits for four modes of EC\,20058$-$5234.}
\end{figure}

We were able to build an O$-$C diagram for four pulsation frequencies (Figure \ref{OC}). First we note that each has a parabolic and sinusoidal component. The parabolic contribution to an O$-$C is indicative of a slow linear change in period. If we fit each O$-$C with a parabola + sinusoid we find that the four O$-$C diagrams can be split into two distinct sets. Two of the O$-$Cs show a positive $\dot{\Pi}$ and a sinusoid with a relative phase of 0 and the other two O$-$Cs reveal a negative $\dot{\Pi}$ and a sinusoid with relative phase $\pi$. The four have a sinusoidal contribution with different amplitudes but similar period. Finally we fit all four O$-$Cs simultaneously with independent parabolas and with sinusoids of identical period and phase. The functional form of our fit to the $i{\rm th}$ O$-$C is 
\begin{equation} \label{model}
Y_{i} = Y_{0i} + \frac{1}{2} \dot{\Pi_i} \Pi_i t^2 + A_{i} sin(\frac{2\pi}{P}t-\phi) 
\end{equation}

The data and the fit is shown in Figure~\ref{OC}. The resultant fit parameters are summarized in Table 1.

\begin{table} \label{tbl1}
\begin{tabular}{lrrrrrrrrr}
\hline
  & \tablehead{1}{r}{b}{$\Pi$}  
  & \tablehead{1}{r}{b}{$A_i$}
  & \tablehead{1}{r}{b}{$\sigma_{A_i}$}
  & \tablehead{1}{r}{b}{$\dot{\Pi_i}$ }
  & \tablehead{1}{r}{b}{$\sigma_{\dot{\Pi_i}}$}
  & \tablehead{1}{r}{b}{$P_i$}
  & \tablehead{1}{r}{b}{$\sigma_{P_i}$}
  & \tablehead{1}{r}{b}{$\phi$}
  & \tablehead{1}{r}{b}{$\sigma_{\phi}$}   \\
  
\hline
& 204.5891 & 10 & 10 & -1 & 2 & 11 & 1 & 2.3 & .1\\
& 256.8549 & -30 & 10 & 4 & 2 & " & " & " & "\\
& 280.9777 & 70 & 10 & -28 & 4 & " & " & " & "\\
& 333.4759 & -110 & 20 & 48 & 7 & " & " & " & "\\
\hline
\end{tabular}
\caption{Model fit parameters. See equation \ref{model}. All units for $\dot{\Pi_i}$ and $\sigma_{\dot{\Pi_i}}$ are $10^{-13}$}
\label{tab:a}
\end{table}

\section{The Effect on Timing Based Planet Searches}
We have discovered sinusoidal behavior in pulsating WD O$-$Cs that are NOT due to a planetary companion. It has been assumed that variations of this nature could only be the result of a planet + WD system. If only one of these four modes had been observed, it is very possible that the topic of this paper would be the first planet detection around a WD. Even more alarming is that the O$-$C diagrams have sinusoidal variations with correlated phase and period. It seems plausible that two modes behaving in this way could conspire to mimic a planet + WD system. This does not eliminate the possibility of a WD planet detection with this method, but demonstrates that a sinusoidal variation in the O$-$C of a single mode is inconclusive. An important question is if DAV and sdBV pulsators also experience the same O$-$C variations. Qualitatively, DAVs and DBVs are very similar. It would seem likely that these types of variations could exist in the DAVs as well. The sdBVs are somewhat different from the WD pulsators, but until the physical processes driving these variations are understood we cannot exclude the possibility of this behavior.


\begin{theacknowledgments}
JD, JLP, and HLS would like to thank the \emph{The Crystal Trust} for financial support.
\end{theacknowledgments}

\bibliographystyle{aipproc}   


\end{document}